\documentclass[english,aps,prd,prstper,reprint,showpacs,titlepage,longbibliography]{revtex4-1}   

\usepackage[T1]{fontenc}	
\usepackage[latin9]{inputenc}	
\usepackage{geometry}		
\usepackage{graphicx}
\usepackage[above,below]{placeins}	
\usepackage{times}

\usepackage{hyperref}  
\hypersetup{colorlinks=true,urlcolor=blue,citecolor=blue,linkcolor=blue}   
\usepackage{enumitem}          
\setlist{nosep}                 

\usepackage{booktabs,tabularx}
\usepackage{color, colortbl}
\definecolor{Gray}{gray}{0.9}
\usepackage{textgreek}
\begin{document}

\begin{titlepage}

  \title{Sense of agency, gender, and students' perception in open-ended physics labs}

  \author{Z. Yasemin Kalender}
    \author{Martin Stein}
    \author{N. G. Holmes}
  \affiliation{Laboratory of Atomic and Solid State Physics, Department of Physics, Cornell University, Ithaca, NY, 14850}  


  \begin{abstract}
    \textbf{Abstract}. Instructional physics labs are critical junctures for many STEM majors to develop an understanding of experimentation in the sciences. Students can acquire useful experimental skills and grow their identities as scientists. However, many traditionally-instructed labs do not necessarily involve authentic physics experimentation features in their curricula. Recent research calls for a reformation in undergraduate labs to incorporate more student agency and choice in the learning processes. In our institution, we have adopted open-ended lab teaching in the introductory physics courses. By using reformed curricula that provide higher student agency, we analyzed approximately 100 students in the introductory-level lab courses to examine their views towards the open-ended physics labs. Between the start and the end of the semester, we found a statistically significant shift in students' perceptions about the agency afforded in lab activities. We also examined students' responses to "Which lab unit was your favorite and why?". The analysis showed that majority of the students preferred  Project Lab, which had the highest student agency and coding analysis showed that "freedom" was the most frequent response for students' reason for picking Project Lab. Finally, we also examined student views across gender and found no significant gender effect on students' sense of agency. 
\clearpage    
  \end{abstract}

  \maketitle
\end{titlepage}

\section{Introduction}

    Scientific thinking is any instance of purposeful thinking that has the objective of enhancing the seeker's knowledge~\cite{kuhn}, and involves many independent and critically engaged decision making processes. Scientific decisions may relate to building models, designing experiments, interpreting results, or communicating the ideas~\cite{AAPT}, among others. To make such decisions, 
    agency plays a critical role, where agency is defined here as the capacity to guide one's actions towards achieving a goal~\cite{Bandura1989, Bandura1996}. 
    To learn to think like a scientist, students need to learn how to make such decisions using their agency.

    
     Undergraduate labs offer a great opportunity to gain practice with such decision making. 
    However, the curricula of many physics lab courses do not satisfy these goals, because they are highly-structured, procedural, and content-reinforcing \cite{PCAST}. 
    Therefore, in recent years, there has been a growing recognition for the need to reform undergraduate physics curricula to adopt more open-ended and student-centered tasks and activities.
    
    Open-ended (i.e., interactive or constructive) lab instruction can offer students more choice and agency 
    as they engage with the material. Open-ended 
    labs can benefit students in multiple ways, from developing more expert-like attitudes towards experimentation~\cite{Wilcox2017} to developing positive science identities~\cite{Doucette2020, Carlone2007,Kalender2019}. Supporting students to enact their agency, however, is more than simply removing structure. Students must have the motivation and self-efficacy to take up the agency afforded~\cite{Bandura1989}.
    In the present work, we studied students' beliefs and perspectives about the agency in a physics lab course and compared across different demographics. 

\section{Literature review and research questions} 


In an open-ended lab, 
students have control over what they do and what they find in their experiment. 
Students have more instances to use their agency to become the initiator of their learning processes \cite{Tapal2017}. 
Affording such agency in lab classes can better engage students in activities that interest them \cite{Scarmadilia2006} and enhance their learning of science practices and career outcomes \cite{Zeiser2018}. Students who perceive themselves as having agency will have more ownership in their learning~\cite{Hanauer2014}. Students with a sense of ownership  will become more invested, engaged, and persist in their work
~\cite{Hanauer2016, Dounas-frazer2017}. 



Supporting agency, however, does not mean full autonomy on students' part~\cite{Bandura1989}. In science education, the goal is not simply for students to make decisions -- the goal is for them to learn to make decisions in line with scientific practice~\cite{Ford2015}. Thus, students should be supported in making 
their own decisions in open-ended course work ~\cite{Holmes2020}. The literature on social learning suggests using teaching processes such as modeling, coaching, scaffolding, and then slowly fading 
support \cite{Cognitive1987}. As students develop higher-order skills and learning activities become more complex, students transition from novices to more independent and expert-level learners ~\cite{Ericsson1993TheRO,chi1981} and they become more autonomous (i.e., higher agency) scientists. This apparent structure may seem antagonistic to students' agency. Given that open-ended labs are rarely fully `open', in what ways are students aware of the agency afforded to them in labs?



Agency in a lab is also 
influenced by the learning environment and social-classroom dynamics and hierarchies \cite{Schenkel}, with such dynamics and hierarchies being imported from the wider cultural context in sciences 
~\cite{Seymour1997}. If one student takes charge in a group, other students might not perceive the agency afforded to them~\cite{Archibeque2017,Archibeque2018}.
In particular, members of marginalized groups might perceive themselves as having less agency to participate in tasks or group discourses because of their peers.  
For example, in physics, researchers have 
found gendered patterns in students' roles 
in lab activities such that female students are assigned to "disempowering" roles
~\cite{Danielsson2012}. The accumulation of many such interactions may impact male and female students' sense of agency differently.


Finally, a critical question to the discussion of agency is whether students value and recognize the benefit of provided agency in physics labs. On one hand, positive attitudes toward agency in physics can influence students' reactions to challenges and setbacks 
\cite{Zeiser2018,dweck2008mindset,elliott1988goals} and students who view agency as positive can become more active-seekers of new knowledge and feel more positively toward self-learning strategies~\cite{Bandura1989}. On the other hand, prior literature shows that students in active-learning classes often feel less positive about and see less value in self-guided learning compared to traditional learning \cite{Cavanagh2016}, despite the fact that they gain more knowledge in the former \cite{Deslauriers2019}.


In this study, we focused on the understudied topic of students' sense of agency in open-ended physics lab courses. We examined three primary research questions.
\begin{itemize}
    \item \textit{RQ1: How do open-ended labs impact students' sense of agency?} While prior literature emphasizes the importance of agency, there is minimal work on how to actually promote students' sense of agency. To address this, we examined whether open-ended labs can be one method of increasing students' sense of agency in physics. 
    \item \textit{RQ2: How does students' sense of agency in these labs interact with gender?} As noted above, student demographics can play an important role in how students engage in physics labs activities. 
    To unpack these complexities in the role of agency in physics labs, we tested for equity of parity (i.e., equity of outcomes) \cite{Rodriquez2012,Vandusen2019,Burkholder2020} to examine whether men and women perceive labs differently at the beginning and end of the instruction.
    \item \textit{RQ3: To what extent do students value having agency in open-ended physics lab?} During a lab course, different assignments will vary in the amount of agency afforded to students. This provides an opportunity to test how students perceive agency by asking for their preferred assignments and the reasoning for those preferences. 
\end{itemize}

\section{Methods}

\subsection{Participants and lab context}

The participants were students in the first semester calculus-based honors physics course sequence at a large research university, who mostly intended to major in physics. There were 88 and 66 students who took the pre and post-test. Students self-identified their gender on the survey and approximately 30\% of the classroom identified as female, one student 
preferred not to disclose their gender, and the remaining students identified as male.  
Labs were combined with the main lecture 
course and taught by three graduate teaching assistants. Each lab session was two hours long and each lab unit typically spanned two sessions. 
In a 15-week semester, there were 9 weeks of lab. Students conducted their experiments in groups of 2-3 students. Between each lab unit, students picked different lab partners. Each unit focused on one or multiple goals such as Model Testing, Ethics, or Model Extending. Afforded student agency in each lab unit (Lab 1, Lab 2, Lab 3) was 
gradually increased towards a student-guided "Project Lab" unit. In this final Project Lab unit, students picked their own research question and topic and designed the whole investigation. 
In the final lab session, students presented their project results to the whole classroom.
More detailed information about each unit can be found on PhysPort \cite{PhysPort}. 


\subsection{Survey development and validation}

\begin{table*}[htbp]
  \caption{SoA items and their factor loadings and \textit{p}-values are shown below. \label{tab2}}
  \begin{ruledtabular}
    \begin{tabular}{lcc}
      \textbf{Items} & \textbf{Factor loadings} & \textbf{\textit{p}-values} \\
      \hline
     I am in control of setting the goals for the experiments. &  0.859 & < 0.001  \\ 
     I have the freedom to design and conduct the best possible experiment to attain my goals. & 0.860 & < 0.001 \\
     I am in control of choosing the appropriate analysis tools
     to evaluate experimental outcomes. & 0.733 & < 0.001\\
     I am in control of doing interesting experiments in a physics lab. & 0.723 & < 0.001
    \end{tabular}
  \end{ruledtabular}
\end{table*}

We administered a short survey 
in the first and last lab session to measure students' attitudes towards experimentation in physics. 
We developed 
Sense of Agency (SoA) survey items 
to measure students' beliefs about the agency afforded in a physics lab context, adapted from an existing survey~\cite{Tapal2017}. 
We 
conducted interviews with three undergraduate and two graduate students to ensure that students interpret the items in the way we intend. These initial student interviews helped us modify the SoA items. The SoA items used 
a Likert Scale and they were chosen to measure different aspects of physics labs such as setting goals, designing and conducting experiments, and choosing analysis tools. We also conducted tests of internal consistency and construct validity by using Cronbach-alpha \cite{Cronbach1955}. The inter-item reliability between SoA questions was 0.87 in the pre-test scores, 0.89 in the post-test, which were considered "good" \cite{Cronbach1951}. The survey included 
additional constructs (e.g., self-efficacy, mindset), and initial factor analysis in R~\cite{RCoreTeam} showed separability between constructs. 
The fit parameters for confirmatory factor analysis were 
CFI = 0.928, TLI = 0.915, RMSEA =  0.078, SRMR = 0.066, which 
are in the range of acceptable model fit \cite{Hooper2008}. Based on the reliability results, among five of the SoA items, four of them met the psychometric standards so we used in our analysis. 
The items and their factor loading values are given in Table \ref{tab2}.

\par In the post attitude survey, we included an additional open-ended question about students' favorite lab unit 
and asked for a short explanation of why they liked that particular lab unit. 
We also interviewed students later in the semester to better understand their experiences in these labs. We are not doing a formal analysis of these data here but will present some student quotes in the discussion section that are relevant to the current study.

\subsection{Analysis}

\subsubsection{Students' Sense of Agency}

In order to answer RQ1 and RQ2, we performed 
$t$-tests and calculated effect sizes (i.e., Cohen's \textit{d} \cite{Cohen1988}) and confidence intervals on whether students' sense of agency changed from pre to post survey, and whether the scores differed between male and female students at either time point. We use alpha value of 0.05 as a threshold for statistical significance level.


\subsubsection{Coding for "Favorite Lab"}

To answer RQ3, we analyzed the open-ended post survey question regarding students' favorite lab and generated a word cloud by using a library in R \cite{RCoreTeam}. The word cloud serves only to provide a qualitative description of students' responses. We emergently and iteratively created a coding scheme from students' open-ended responses. 
The final coding scheme is presented in Table \ref{tab_codes}. Note that 
multiple codes could apply to 
a single response. Two authors independently coded all of the items with the final coding scheme. An inter-rater reliability showed "good" or "excellent" agreement between the coders. Table \ref{tab_codes} shows percentage of agreement and Cohen's Kappa values. All disagreements were discussed to reach consensus.


\begin{table*}[htbp]
  \caption{Coding table with code names, their definition, and an example sentence from students. Percent agreement and Cohen's Kappa values are also given for an inter-reliability to measure between the coders. A particular student response can be labeled with multiple codes. \label{tab_codes}}
  \setlength\tabcolsep{5pt}
    \begin{tabularx}{\textwidth}{p{2cm}XXcc} 
\toprule
      \textbf{Code name} & \textbf{Definition} & \textbf{Example} & \textbf{Percent} & \textbf{Cohen's}  \\
      & & & \textbf{Agreement} & \textbf{Kappa} \\
      \hline
    
     Learning concepts & Experiment helped learning theory and/or concepts & "Allowed me to study aspects of Hooke's law that I did know before" & 94.8 & 0.69 \\
     Interest & Experiment was fun / interesting / enjoyable / engaging  & "I was genuinely excited by the material..." & 92.2 & 0.79\\
 
     Logistics & Experiment was short / organized (or not) / clear (or not) / structured (or not) & "It was more organized ..." & 97.4 & 0.79 \\
 
     Epistemological belief & Experiment provoked the most critical thinking, had unexpected results & "That unit proved to me that the lab is also about disconfirming predictions for an experiment...." & 93.5 & 0.67 \\

     Group work & Experiment was the favorite because of group mates & "Good group, worked well together..." & 97.4 & 0.82 \\
  
     Freedom & Flexibility, creativity regarding the research questions, design or in general the unit itself & "Because in project lab session, we have the freedom to explore the subject we are interested in and be able to design our own experiment." &98.7 & 0.97 \\ 
     Other & No response/"No lab in high school"/too vague to code & "Better than the other ones" &90.0 & 0.76 \\
        \hline\hline
    \end{tabularx}
\end{table*}

\section{Results}

\begin{figure}[ht]
  \includegraphics[width=0.50\textwidth]{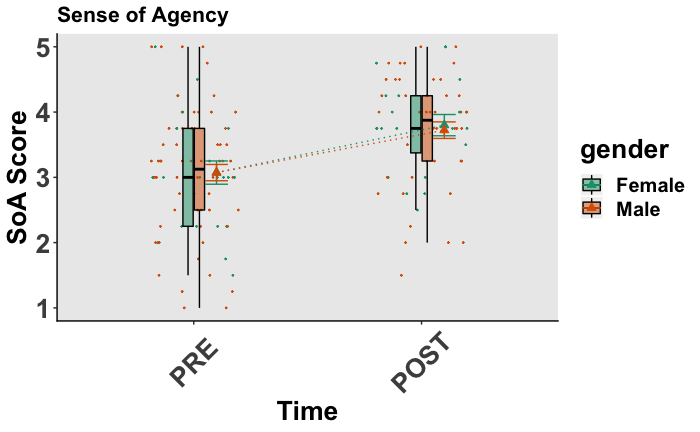}
  \caption{Raincloud plot \cite{Raincloud} presents the shift in students' Sense of Agency scores from pre to post-test by gender. Box plot, error bars, and the distribution of students' scores are included in the plot. The dotted lines shows the trend in students' agency belief shift by gender. \label{fig:PA}}
\end{figure}

To answer RQ1, we first checked the variance between pre and post scores and found no difference (\textit{p} = 0.213). Then we performed Students' $t$-test where we assumed equal variance based on the previous analysis and found that students' average SoA 
scores 
increased from $M$ = 2.96, $SD$ = 0.98 at pre-test to post-test $M$ = 3.73, $SD$ = 0.84 with $t(56) = 5.29$, \textit{p} < 0.001. The mean difference in SoA between pre and post test is 0.77 with 95\% Confidence Interval [0.48,1.07] (see Fig \ref{fig:PA}). The effect size between pre and post SoA scores is 0.76, which is considered as between "medium" and "large" \cite{Cohen1988}.

To address RQ2, we compared 
SoA scores between female and male students at both pre and post-test separately. We did not find any gender difference in students' SoA either in the pre-test $t$(84) = 0.01, \textit{p} = 0.99 or the post-test $t$(64) = 0.35, \textit{p} = 0.72. 



\begin{figure}[ht]
  \includegraphics[width=0.30\textwidth]{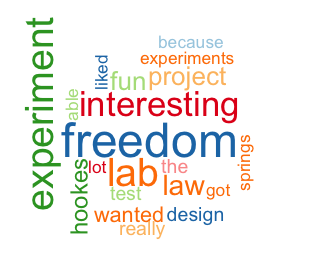}
  \caption{Using rquery.wordcloud() function, word cloud generator shows most frequently used word for students' open-ended responses to why they liked a particular lab unit. "Freedom" is the most frequently written word, i.e., 25\% students wrote "freedom" for why they liked a particular lab unit. \label{fig:wordcloud1}}
\end{figure}
\begin{figure}[ht]
  \includegraphics[width=0.50\textwidth, 
  ]{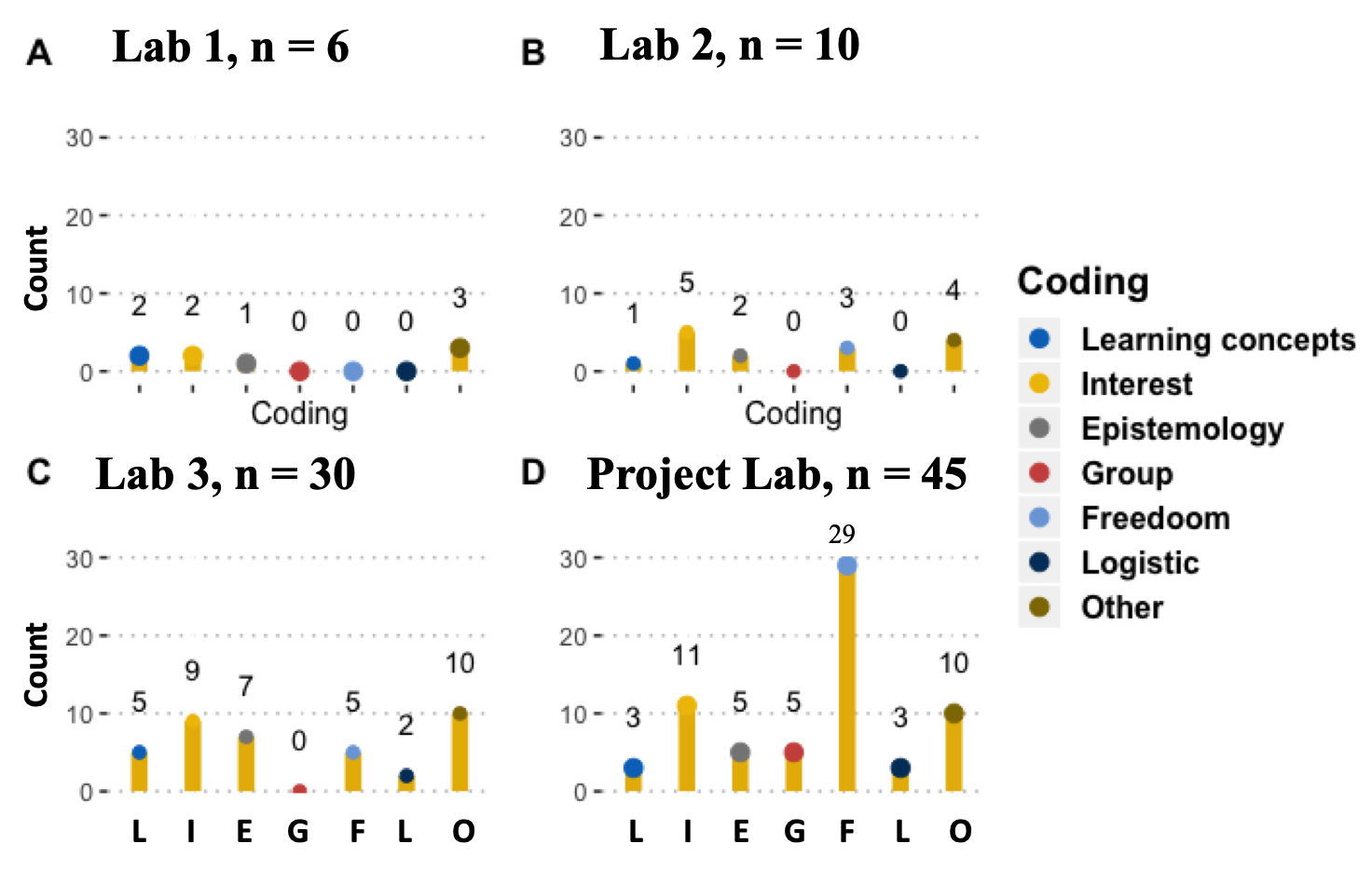}
  \caption{Coding results for students' explanation to their preferred lab unit. There are seven main codes: Learning concept (L), Interest(I), Epistemology (E), Group (G), Freedom (F), Logistic (L) and Other (O). There are four lab units (Lab 1, Lab2, Lab3, and Project Lab) and each unit takes two weeks to complete. Number of students who choose specific lab as their favorite unit is shown with "n". \label{fig:coding1116}}
\end{figure}

When we looked at the distribution of students' favorite lab, we found that the majority of the students (n = 45) chose the Project Lab unit as their favorite, followed by Lab 3 (n = 30). The word cloud analysis presented in Fig. \ref{fig:wordcloud1} indicates that students most frequently stated "freedom" as the reason for why they liked any particular lab unit. Based on our coding analysis, as seen in Figure \ref{fig:coding1116}, the most common explanation given by students who selected the Project Lab as their favorite unit was the freedom that the lab afforded them. "Interest" was the second most common explanation. 
The first two lab units 
were selected by very few students as their favorite lab, 6 and 10 students, respectively. The coding patterns for Lab 1 and Lab 2 were very similar. 






\section{Conclusion and Discussion}

How students take up agency and whether they view agency as positive is critical in physics lab courses \cite{Dounas-frazer2017,Zeiser2018}. We found support that open-ended labs can increase students' perceptions about their agency. Specifically, addressing RQ1, we examined the relationship between students' SoA and time, and found that students indicated having more agency (or a greater sense of agency) by the end of the course. In interviews at the end of the semester, many students referred to the notion of freedom explicitly, such as:
"\textit{There was more freedom for us to choose how to design and carry out an experiment.}" 
Given that prior literature has shown students' attitudes in physics tend to decline as the semester continues \cite{Adams2006}, our findings provide evidence that open-ended labs can elevate students' views towards agency, potentially buffering against these typical declines in motivation. 

In order to address RQ2, we also tested whether students' SoA varied between male and female students. 
Although previous work suggests that female students experience a larger drop in attitudes than male students in physics courses \cite{kalender2019diversity}, we found that both female and male students' sense of agency increased over the semester. Given that male students typically dominate group conversations in physics \cite{Heller1992}, one possibility that could have happened was that male students could have increased in their SoA more than female students. However, we found that there were no gender differences in SoA scores, both at the start and at the end of the semester. This suggests that open-ended labs with student agency and frequent peer collaboration can help both female and male students' motivation.
We note that this analysis focused on the gender as binary, however, we do acknowledge that gender is a fluid, multi-level, and deeply complex construct.

 The design and instruction of these open-ended labs incrementally reduced structure, making room for student agency; 
instruction built towards the student-guided "Project Lab" unit instead of beginning with it. Thus, we examined students' reasoning for their preferred lab in addressing RQ3. We found that students most frequently preferred the Project lab unit, suggesting that they embraced self-directed learning (i.e., freedom, see Fig.\ref{fig:coding1116}). Students' responses to why they liked Project Lab most often include "freedom" in terms of coming up with their own research question, designing experiments, or being able to be creative in their experimentation. Even though prior work found that students resist against self-guided learning \cite{Seidel2013}, our results suggest that this might not always be the case and that the particular method of implementing self-guided learning is crucial. Students may have felt positively, rather than frustrated, about the open-endedness because the student agency was expanded in each unit and students were never fully autonomous. Prior work highlights the importance of balancing structure and open-endedness in teaching via gradual increase in student agency \cite{Cognitive1987}. In an interview at the end of the semester, one student described the balance as follows: "\textit{I definitely think that it is a good idea to have structure. That being said, I created the structure myself and am able to bend it. Maybe designing our labs has helped. That feels like we had ownership.}" 




Classroom activities with hands-on tasks offer opportunities to teach students to be more agentic in their learning processes. The learning environment and interactions with peers during class can enhance or hinder student agency, which can influence their confidence and identity  \cite{Godwin2016}. Particularly, first-year college experiences can have important downstream consequences regarding motivation in specific courses and broader academic (e.g., enrolling in more physics classes) and career (e.g., retention in STEM fields) choices \cite{Godwin2016,Hazari2010}. 
Other research indicates that such agency-enhancing activities facilitate student learning, and the research here indicates that students are likely to perceive that agency positively and without differential impacts by gender.



\acknowledgments{This material is based upon work supported by the National Science Foundation under Grant No. 1836617. We also thank the Cornell Active Learning Initiative for their support. Finally, we thank the Cornell Physics Education Research Lab members for their constructive feedback throughout this study}

\clearpage
\bibliography{Main_text.bib}

\end{document}